\pdfoutput=1
\documentclass[sigconf]{acmart}

\AtBeginDocument{%
  }

\setcopyright{none}
\renewcommand\footnotetextcopyrightpermission[1]{}
\acmConference[preprint]{preprint}{2026}{}
\acmYear{2026}

\usepackage{booktabs}
\usepackage{amsmath}
\usepackage{algorithm}
\usepackage{algpseudocode}
\usepackage{graphicx}
\usepackage{placeins}

\begin{document}

\title{Hierarchical Reinforced Trader (HRT): A Bi-Level Approach for Optimizing Stock Selection and Execution}

\author{Zijie Zhao}
\affiliation{%
  \institution{Massachusetts Institute of Technology}
  \city{Cambridge}
  \state{MA}
  \country{USA}}
\email{zijiezha@mit.edu}

\author{Roy E. Welsch}
\affiliation{%
  \institution{Massachusetts Institute of Technology}
  \city{Cambridge}
  \state{MA}
  \country{USA}}
\email{rwelsch@mit.edu}

\begin{abstract}
Automated equity trading requires converting noisy market and news signals into executable portfolio decisions under risk, turnover, and transaction costs. We propose \textbf{Hierarchical Reinforced Trader (HRT)}, a bi-level reinforcement learning framework for text-aware portfolio management in multi-asset equity markets. HRT separates trading into two coordinated decisions: a factorized sparse \textit{High-Level Controller (HLC)} selects asset-level increase, reduce, or hold directions from compact market and text-derived signals, while a risk-aware \textit{Low-Level Controller (LLC)} converts these directions into feasible portfolio weight adjustments under turnover, drawdown, and text-risk penalties. This decomposition avoids enumerating the full joint action space and makes selection and execution easier to inspect. We evaluate HRT on an open stock-news benchmark with a fixed 89-stock Nasdaq universe, using 2013--2018 for training, 2019 for validation, and 2020--2023 for final out-of-sample testing; the test horizon is restricted to 2020--2023 due to public benchmark data availability under the same timestamp-clean text-aware protocol. Across market-proxy, same-universe portfolio, alpha-only, flat-RL, and hierarchical ablation baselines, HRT delivers the strongest learning-based return--risk--cost trade-off. The full model improves Sharpe from 1.06 for HRT-Base to 1.24, reduces daily turnover from 0.112 to 0.090, and remains robust under transaction-cost stress. These results suggest that separating sparse directional selection from risk-aware execution is an effective way to incorporate market forecasts and text-derived risk signals into portfolio management.
\end{abstract}



\maketitle

\section{Introduction}
\label{sec:intro}

Automated equity trading requires repeated decisions under noisy signals, changing market regimes, and practical trading frictions. Classical portfolio construction, beginning with mean--variance optimization \cite{markowitz1952portfolio}, provides a principled way to balance return and risk, but it is less direct for sequential decision-making with changing signals, transaction costs, and portfolio constraints. A common alternative is to formulate trading as a Markov Decision Process (MDP) \cite{bertsekas2012dynamic} and learn trading policies with reinforcement learning. Deep reinforcement learning (DRL) has been applied to stock trading and portfolio allocation using algorithms such as DDPG, PPO, A2C, TD3, and ensemble strategies \cite{lillicrap2015continuous,schulman2017proximal,mnih2016asynchronous,fujimoto2018addressing,liu2018practical,yang2020deep,liu2021finrl}. These methods are flexible, but applying a single flat policy to a large stock universe remains challenging. In this work, we focus on text-aware portfolio management in multi-asset equity markets, where the goal is not only to forecast returns, but also to translate market and text-derived signals into feasible, cost-aware portfolio rebalancing decisions.

The difficulty is not only forecasting returns. In multi-stock trading, a policy must convert noisy market and text-derived signals into executable portfolio actions. Three issues are particularly important. First, the joint action space grows quickly with the number of assets when a policy simultaneously decides directions and trade sizes. Second, raw market factors and raw text can be high-dimensional and noisy, making end-to-end policy learning unstable and difficult to interpret. Third, profitable directional signals do not automatically translate into good executions: excessive turnover, drawdown, concentration, and transaction costs can eliminate apparent gains. These concerns are especially important in financial RL, where benchmark quality, backtesting overfitting, and data leakage are common sources of fragility \cite{liu2022finrlmeta,liu2024dynamic}.

Financial text and large language models offer useful complementary information, but they also raise a design question. Recent financial language models and stock-news datasets make it possible to extract sentiment, event, and risk-related signals from news \cite{zhang2023instruct,dong2024fnspid}. More recent systems explore LLM-augmented or multimodal trading agents \cite{zhang2024finagent,benhenda2025finrldeepseek}. In this work, we do not use language models as autonomous trading agents. Instead, text is converted into timestamp-aligned structured sentiment and risk signals, and the RL policy remains responsible for trading decisions. This keeps the decision process auditable and allows textual risk information to affect both stock selection and position sizing.

We introduce the \textbf{H}ierarchical \textbf{R}einforced \textbf{T}rader (\textbf{HRT}), a bi-level framework that separates directional stock selection from portfolio execution. HRT decomposes the trading decision into sparse high-level direction selection and risk-aware low-level execution, as illustrated in Figure~\ref{fig:HRT}. The High-Level Controller (HLC) uses a factorized categorical policy to produce sparse increase, reduce, or hold decisions for each asset. This avoids enumerating the full $3^N$ joint direction space while preserving asset-level flexibility. The Low-Level Controller (LLC) receives the HLC directions and current portfolio state, then determines risk-aware portfolio adjustments in weight space. Its reward includes net return, turnover, drawdown, and exposure to text-derived risk. The two controllers are trained with a phased alternating procedure: the HLC first learns usable directional decisions, the LLC then learns constrained execution, and later updates use execution feedback to align high-level selection with portfolio-level outcomes.

We evaluate HRT on the open FinRL-DeepSeek stock-trading benchmark, a FinRL-compatible benchmark constructed from FNSPID financial news and daily market data with precomputed LLM-generated sentiment and risk signals. The model is trained on 2013--2018, validated on 2019, and tested once on 2020--2023, covering the COVID shock and recovery, a bull market, a Fed-tightening drawdown, and a technology-led recovery. The experiments compare HRT with an external market proxy, same-universe portfolio baselines, alpha-only selection, flat RL, and HRT ablations. Under this fixed benchmark protocol, HRT improves the return--risk--cost trade-off among learning-based strategies. Compared with HRT-Base, HRT improves Sharpe from 1.06 to 1.24 and reduces daily turnover from 0.112 to 0.090, while not claiming to dominate every individual risk metric.

Our main contributions are:
\begin{itemize}
    \item We formulate text-aware portfolio management in multi-asset equity markets as a bi-level hierarchical decision problem that separates sparse directional selection from risk-aware portfolio execution.
    \item We introduce a factorized sparse High-Level Controller that produces asset-level increase, reduce, or hold decisions without enumerating the joint $3^N$ direction space.
    \item We design a structured signal interface and risk-aware Low-Level Controller that incorporate market forecasts, benchmark-provided text sentiment, and text-risk signals into constrained portfolio adjustment.
    \item We evaluate HRT on an open stock-news benchmark with same-universe baselines, ablations, and transaction-cost stress tests, showing a stronger learning-based return--risk--cost trade-off.
\end{itemize}

The remainder of the paper is organized as follows. Section~\ref{sec:relatedwork} reviews related work. Section~\ref{sec:methodology} presents the HRT framework. Section~\ref{sec:experiments} describes the benchmark, experimental protocol, and empirical results. Section~\ref{sec:conclusions} discusses limitations and concludes.

\section{Related Work}
\label{sec:relatedwork}

\subsection{Reinforcement Learning and Benchmarks for Portfolio Trading}

Reinforcement learning has been widely studied for automated trading because it naturally models sequential decision-making under changing market states. Early practical DRL trading systems use deterministic actor--critic methods such as DDPG to map portfolio states to continuous trading actions \cite{lillicrap2015continuous,liu2018practical}. Later work evaluates PPO, A2C, DDPG, and ensemble policies under the FinRL framework \cite{schulman2017proximal,mnih2016asynchronous,yang2020deep,liu2021finrl}. TD3 improves deterministic actor--critic learning by addressing value overestimation through clipped double-Q learning and delayed policy updates \cite{fujimoto2018addressing}. These methods provide useful baselines, but they typically learn a flat policy that directly selects portfolio actions, making it difficult to separate stock selection from execution constraints in large universes.

Recent financial RL benchmarks emphasize reproducibility, market environment design, and realistic data splits \cite{liu2022finrlmeta,liu2024dynamic}. These benchmarks are important because trading agents are particularly vulnerable to overfitting, look-ahead bias, and inconsistent evaluation protocols. HRT follows this benchmark-oriented approach: model selection is performed on a validation period, and the test period is used only for final out-of-sample evaluation.

\subsection{Hierarchical Reinforcement Learning for Trading}

Hierarchical reinforcement learning decomposes long-horizon decision problems into higher-level and lower-level controllers \cite{pateria2021hierarchical}. In financial applications, hierarchy has been used to combine portfolio selection and order execution or to structure high-frequency trading decisions. For example, hierarchical portfolio and execution models separate allocation from trade implementation \cite{wang2020deep}, and high-frequency systems such as EarnHFT use hierarchical policies to improve trading efficiency \cite{qin2024earnhft}. Select-and-trade approaches further show that hierarchical structure can help separate asset selection from transaction decisions \cite{han2023select}. HRT differs from these works by focusing on daily multi-stock equity trading with structured market/text signals, factorized sparse direction selection, and a risk-aware execution controller.

\subsection{Financial Text and LLM-Derived Trading Signals}

Financial news and textual disclosures can provide information not captured by price and volume alone. Sentiment-aware DRL and knowledge-enhanced trading systems incorporate text-derived features to improve portfolio decisions \cite{koratamaddi2021market,nan2022sentiment}. More recent financial language models, such as FinGPT, show that instruction-tuned models can extract financial sentiment from text \cite{zhang2023instruct}. FNSPID provides a large-scale time-aligned financial news and price dataset, enabling open evaluation of stock-news trading systems \cite{dong2024fnspid}. Recent LLM-augmented trading systems further explore multimodal financial agents or risk-sensitive RL with LLM-generated signals \cite{zhang2024finagent,benhenda2025finrldeepseek}. HRT adopts a constrained role for language models: text is converted into structured sentiment and risk signals, and the hierarchical RL policy remains responsible for trading decisions.

\section{Methodology}
\label{sec:methodology}

\subsection{Overview}
\label{sec:method_overview}

\begin{figure*}[t]
\centering
\includegraphics[width=\textwidth]{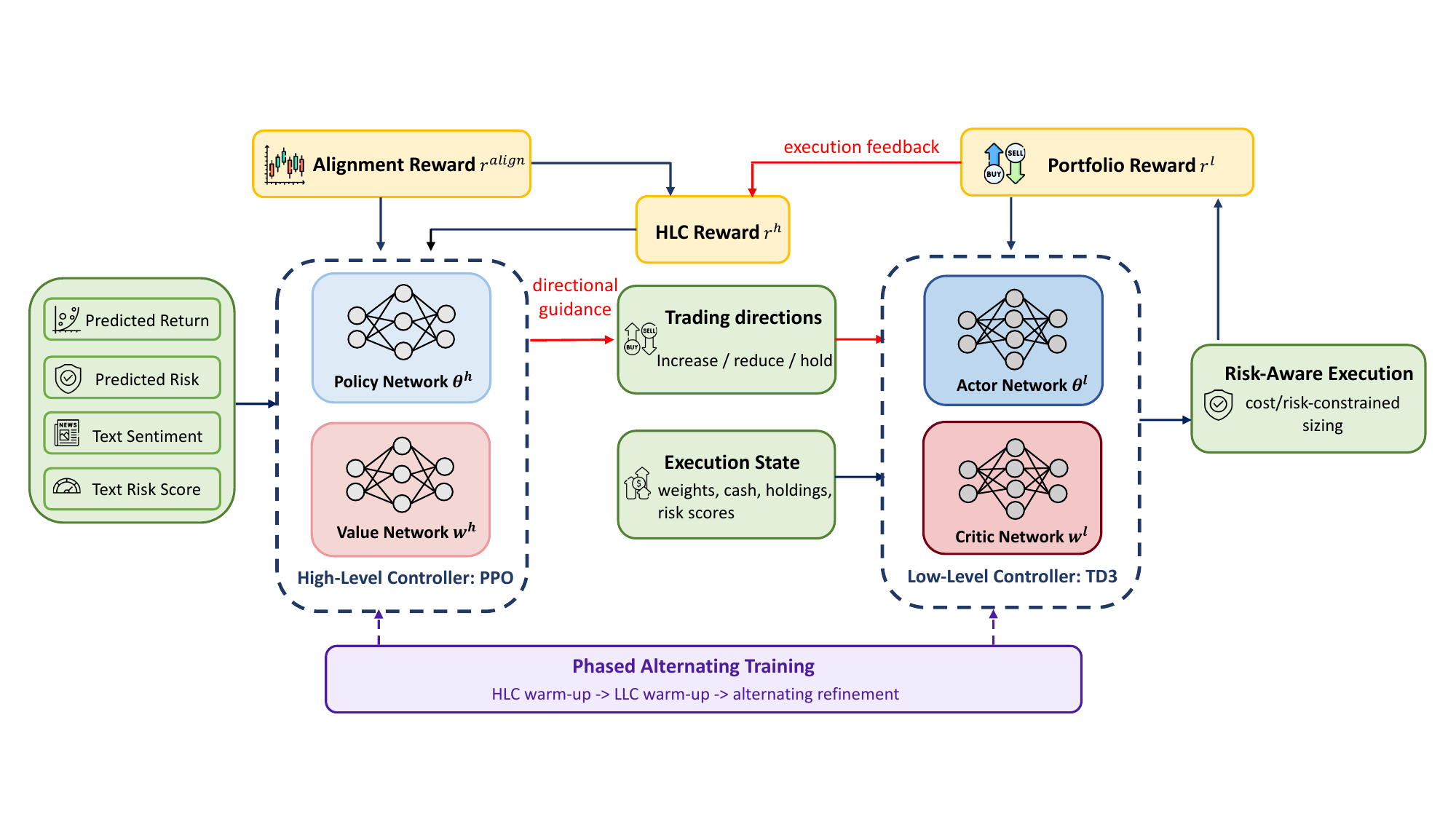}
\caption{Overview of the Hierarchical Reinforced Trader (HRT). The HLC selects sparse trading directions from structured financial signals, while the LLC converts these directions into cost- and risk-aware portfolio adjustments. Red arrows indicate the two key interactions between the controllers: directional guidance from the HLC to the LLC and execution feedback from the LLC to the HLC.}
\Description{A framework diagram showing structured financial signals entering a High-Level Controller, trading directions and portfolio state entering a Low-Level Controller, and feedback from portfolio reward to high-level reward.}
\label{fig:HRT}
\end{figure*}

The Hierarchical Reinforced Trader (HRT) is a risk-aware hierarchical reinforcement learning framework for equity trading. As shown in Figure~\ref{fig:HRT}, HRT consists of two components with distinct responsibilities. The HLC forms sparse directional views by selecting whether each asset should be increased, reduced, or held. The LLC then translates these directions into portfolio weight adjustments subject to trading and risk constraints. This decomposition reflects a practical distinction in portfolio management: alpha signals form directional views, whereas execution must also account for costs, turnover, current holdings, drawdown, and risk exposure.

Compared with a flat trading policy that directly maps a high-dimensional state to trading quantities for all assets, HRT is built around three design principles. First, the HLC uses a factorized sparse policy, which parameterizes stock-level directional decisions without enumerating the full joint action space. Second, the policy consumes compact structured financial signals rather than raw market factors or raw text. Third, the LLC performs risk-aware sizing: it follows the HLC direction mask while penalizing excessive turnover, drawdown, and exposure to text-derived risk.

\subsection{Structured Financial Signal Interface}
\label{sec:signal_interface}

HRT is agnostic to the specific alpha model or language model used upstream. For asset $i$ on day $t$, the HLC receives a compact state
\begin{equation}
    s^h_{i,t} =
    [\bar w_{i,t},\, \hat r_{i,t},\, \hat \sigma_{i,t},\, u_{i,t},\, \rho_{i,t}],
    \label{eq:hlc_state}
\end{equation}
where $\bar w_{i,t}$ is the current pre-trade portfolio weight, $\hat r_{i,t}$ is a predicted forward return, $\hat \sigma_{i,t}$ is a risk proxy, $u_{i,t}$ is a structured textual sentiment score, and $\rho_{i,t}$ is a structured textual risk score. Including the current holding makes the reduce action meaningful in a long-only portfolio: reducing a zero-weight asset is mapped to no trade by the execution layer. We use asset-level notation when defining the HLC input; vector quantities such as $\hat r_t$ and $\rho_t$ later denote the corresponding signals across the tradable universe.

In our experiments, the market return forecast is produced by a LightGBM model \cite{ke2017lightgbm} trained on OHLCV-derived technical indicators and Qlib-style factors \cite{yang2020qlib} to predict the next tradable-period return. The risk proxy $\hat\sigma_{i,t}$ is computed as the 20-day rolling realized volatility using only past returns. For textual inputs, we use the processed benchmark's precomputed LLM-generated sentiment and risk signals, which are derived from timestamp-aligned financial news rather than generated online during RL training. These signals are already aligned with stock-day records in the benchmark; before RL training, we normalize them using training-period statistics and cache the resulting policy inputs. No LLM fine-tuning or online LLM querying is performed during RL training or evaluation.

For each trading decision, the policy uses benchmark-provided market and text-derived signals available before the close-of-day decision cutoff on day $t$. The portfolio is rebalanced at the next tradable open, and realized returns are used only as ex-post rewards or evaluation targets. When necessary, after-cutoff or non-trading-day records are rolled forward to the next tradable decision date. Thus, text-derived signals are used as decision-time inputs rather than ex-post labels.

\subsection{Factorized Sparse High-Level Controller}
\label{sec:factorized_hlc}

The HLC selects stock-level trading directions. For asset $i$, the high-level action is $a^h_{i,t}\in\{-1,0,1\}$, where $1$ denotes increasing the position, $-1$ denotes reducing the position, and $0$ denotes holding it unchanged. Rather than treating the HLC decision as an unstructured joint action over $3^N$ possible direction vectors, we use a factorized categorical policy
\begin{equation}
    \pi^h_{\theta}(a^h_t\mid s^h_t)
    =
    \prod_{i=1}^{N}
    \pi^h_{\theta}(a^h_{i,t}\mid s^h_{i,t}).
    \label{eq:factorized_policy}
\end{equation}
The policy is implemented with shared network parameters and asset-level directional logits. Thus, the HLC evaluates $O(N)$ categorical distributions and produces $3N$ logits per day, rather than enumerating a joint $3^N$ action space. Sparse activation further reduces the LLC sizing problem from all $N$ assets to the active set of assets with nonzero executed adjustments.

The HLC is trained with PPO \cite{schulman2017proximal}. Sparsity is induced through the hold action, a validation-selected confidence filter, and an activation penalty. Its reward combines directional alignment with downstream execution feedback:
\begin{equation}
    r^h_t
    =
    \alpha_t \widetilde r^{align}_t
    +
    (1-\alpha_t)\widetilde r^l_t
    -
    \lambda_{act}\frac{|\mathcal{A}^{exec}_t|}{N}.
    \label{eq:hlc_reward}
\end{equation}
Here $\mathcal{A}^{exec}_t$ is the set of assets with nonzero executed weight changes, and the tildes denote rewards normalized using training-period running statistics. The alignment component is computed only over executed changes; if the HLC produces an all-hold decision or a proposed reduce action is infeasible because the current holding is zero, the corresponding alignment contribution is zero. Normalizing both the alignment term and LLC feedback before mixing gives $\alpha_t=\alpha_0\exp(-\lambda t)$ a stable interpretation as training shifts from directional correctness toward portfolio-level outcomes. Next-day realized returns are used only to compute ex-post training rewards and evaluation metrics. They are never included in the HLC or LLC state observed at decision time.

\subsection{Risk-Aware Low-Level Controller}
\label{sec:risk_aware_llc}

Given the HLC directions, the LLC determines the portfolio adjustment. We formulate the LLC action in weight space rather than raw share counts, because portfolio weights make risk and turnover constraints explicit. Let $\bar w_t$ denote the current pre-trade portfolio weights, $V_t$ the portfolio value, and $d_t=(V^{peak}_t-V_t)/V^{peak}_t$ the current drawdown state. The LLC observes
\begin{equation}
    s^l_t=
    [\bar w_t,\, b_t/V_t,\, d_t,\, a^h_t,\, \hat r_t,\, \hat\sigma_t,\, \rho_t],
    \label{eq:llc_state}
\end{equation}
where $b_t$ is cash and $\rho_t$ is the vector of normalized text-derived risk scores.

The TD3 actor first proposes a raw weight adjustment. Before execution, HRT applies a deterministic feasibility layer:
\begin{equation}
\begin{aligned}
    \widetilde{\Delta w}_t &= \mu_{\phi}(s^l_t),\\
    \Delta w^{exec}_t &= \mathcal{F}(\widetilde{\Delta w}_t,a^h_t,\bar w_t),\\
    w_t &= \bar w_t + \Delta w^{exec}_t.
\end{aligned}
\label{eq:llc_action_chain}
\end{equation}
The feasibility layer is intentionally simple. It first clips each adjustment to match the HLC direction: increase permits only nonnegative changes, reduce permits only reductions down to zero, and hold sets the adjustment to zero. It then clips weights to the single-name limit and scales only the active positive adjustments when buy demand exceeds available cash or the daily turnover budget, so that hold assets remain unchanged and the HLC direction constraints are preserved. The executed feasible adjustment, rather than the raw actor output, is stored in the replay buffer and used by the TD3 critic. During actor updates, the actor objective is evaluated through the same feasibility layer, i.e., on 
$F(\mu_\phi(s_t^l), a_t^h, \bar{w}_t)$ rather than the unconstrained raw adjustment. This keeps both the critic and actor aligned with the action actually executable in the trading environment while avoiding an additional optimization layer.

The LLC reward optimizes net, risk-aware execution:
\begin{equation}
    r^l_t
    =
    R^{net}_t
    -
    \lambda_{turn}\mathrm{Turnover}_t
    -
    \lambda_{dd}d_{t+1}
    -
    \lambda_{risk}\sum_i w_{i,t}\rho_{i,t}.
    \label{eq:llc_reward}
\end{equation}
Here $R^{net}_t=(V_{t+1}-V_t)/V_t$ is the portfolio return after transaction costs have been deducted by the environment, $d_{t+1}$ is the post-trade drawdown, and the final term penalizes exposure to assets with high text-derived risk. This reward does not assume that lower text-risk exposure always improves raw return; instead, it encourages the execution layer to trade off return, cost, turnover, and downside control.

We implement the LLC with a TD3-style actor--critic controller \cite{fujimoto2018addressing}. TD3 is suitable for this continuous-control execution problem because clipped double-Q learning and delayed policy updates reduce overestimation in deterministic actor--critic training. Baseline variants are described in Section~\ref{sec:experiments}.

\subsection{Phased Alternating Training}
\label{sec:phased_training}

The two controllers are trained with a phased alternating procedure. The purpose is not to add another optimization heuristic, but to preserve the interpretation of the hierarchy: the HLC should first learn usable directional decisions, and the LLC should then learn how to size these decisions under portfolio constraints.

\begin{algorithm}[H]
\caption{Training Procedure for HRT}
\label{alg:hrt_training}
\begin{algorithmic}[1]
\State Prepare market forecasts $(\hat r,\hat\sigma)$ from benchmark market data and load benchmark-provided text sentiment and risk signals $(u,\rho)$ on the training period.
\State Normalize text-risk scores using training-period statistics.
\State Initialize the HLC policy $\pi^h_{\theta}$ and the LLC actor--critic networks.
\State \textbf{Phase 1: HLC warm-up.} Train $\pi^h_{\theta}$ using PPO with normalized directional-alignment rewards; LLC feedback is not used in this phase.
\State \textbf{Phase 2: LLC warm-up.} Freeze $\pi^h_{\theta}$ and train the LLC to execute the HLC directions using the risk-aware reward in Eq.~\eqref{eq:llc_reward}.
\State \textbf{Phase 3: Alternating refinement.} Alternately update HLC and LLC. In the HLC reward, decay $\alpha_t$ so that LLC feedback gradually receives more weight. After HLC updates, new trajectories are collected under the updated hierarchy.
\State Select sparse thresholds, risk penalties, and checkpoints using validation-period risk-adjusted performance.
\end{algorithmic}
\end{algorithm}

During alternating refinement, LLC transitions store the HLC action that produced the execution decision. The replay buffer stores the feasible portfolio adjustment actually executed by the environment, rather than the unconstrained actor output, because rewards are generated after enforcing trading constraints. All model selection decisions, including sparse confidence thresholds, risk penalty coefficients, and checkpoints, are made using the 2019 validation period only. The 2020--2023 test period is used once for final out-of-sample evaluation.

\subsection{Computational Considerations}
\label{sec:computational_considerations}

The factorized HLC requires $O(N)$ directional evaluations per day rather than enumerating a joint $3^N$ action space. Sparse activation further reduces the number of assets passed to the LLC from $N$ to $|\mathcal{A}^{exec}_t|$. Market forecasts are computed offline before trading decisions, and benchmark-provided text signals are loaded as structured stock-day inputs. The online policy consumes only compact forecasts, sentiment scores, and risk scores. This design keeps the training cost comparable to standard financial RL baselines while making the framework suitable for larger open equity universes.

\section{Experiments}
\label{sec:experiments}

We design the experiments to evaluate whether HRT improves the conversion of compact market and text-derived signals into executable portfolio decisions. The emphasis is not on maximizing raw return at all costs, but on improving the trade-off among return, downside risk, turnover, and transaction costs. We therefore evaluate HRT along three dimensions: out-of-sample performance, component-level ablations, and robustness to transaction-cost stress. Unless otherwise stated, deterministic baselines are computed once, and stochastic learning-based methods are averaged over five random seeds.

\subsection{Benchmark and Experimental Setup}
\label{sec:benchmark_setup}

We evaluate HRT on the open FinRL-DeepSeek stock-trading benchmark, a FinRL-compatible benchmark constructed from FNSPID financial news and daily market data with precomputed LLM-generated sentiment and risk signals \cite{benhenda2025finrldeepseek, dong2024fnspid,liu2021finrl}. The tradable universe is the filtered 89-stock Nasdaq universe released with the benchmark. We follow this fixed universe for reproducibility; it should not be interpreted as a point-in-time dynamic tradable universe, and survivorship effects may not be fully removed. The model is trained on 2013--2018, validated on 2019, and tested on 2020--2023. We restrict the main evaluation to 2020--2023 because extending the same text-aware protocol beyond the public benchmark horizon would require constructing an additional timestamp-clean news-signal cache.

\begin{table}[t]
\centering
\caption{Benchmark protocol. QQQ is used only as an external Nasdaq-100 market proxy; all portfolio baselines and HRT variants are evaluated on the same tradable stock universe.}
\label{tab:benchmark_protocol}
\small
\resizebox{\columnwidth}{!}{%
\begin{tabular}{ll}
\toprule
\textbf{Item} & \textbf{Setting} \\
\midrule
Universe & Fixed 89-stock Nasdaq universe \\
Train / Val / Test & 2013--2018 / 2019 / 2020--2023 \\
Market signals & OHLCV factors, technical indicators, $\hat r$, realized-volatility proxy \\
Text signals & Timestamp-aligned stock-day sentiment and risk scores \\
Trading protocol & Features/news through close $t$; after-cutoff news rolled forward \\
Execution & Daily weight-based rebalance at next tradable open $(t+1)$ \\
Base cost & 10 bps per traded notional \\
Market reference & QQQ, external Nasdaq-100 proxy only \\
Seeds & 5 seeds for stochastic learning-based methods \\
Model selection & Validation-only selection on 2019 \\
\bottomrule
\end{tabular}%
}
\end{table}

For each rebalancing date, we use the benchmark-provided time-aligned market, news, and LLM-generated text signals available before the close-of-day decision cutoff on day $t$. The portfolio is rebalanced at the next tradable open, and P\&L is measured until the next scheduled rebalance. News or signal updates released after the cutoff, including after-hours and non-trading-day records, are assigned to the next tradable decision date. This protocol is applied consistently to training, validation, and testing. The text-derived sentiment score $u_{i,t}$ and risk score $\rho_{i,t}$ are benchmark-provided stock-day signals that are normalized using training-period statistics and cached before RL training.

\paragraph{Baselines.}
We compare HRT with three groups of baselines. First, we report QQQ as an external Nasdaq-100 market proxy to contextualize the market environment. Second, we compute same-universe portfolio baselines. Equal Weight rebalances the benchmark universe uniformly. Minimum Variance uses a rolling covariance estimate with long-only and single-name weight constraints. Momentum Top-$K$ ranks assets by past 20-day return and holds the top-$K$ names with equal weights. Alpha Top-$K$ ranks assets by the same predicted forward return $\hat r_{i,t}$ used by HRT and holds the top-$K$ names without reinforcement learning. Third, we compare with Flat TD3 and HRT variants.

All same-universe portfolio strategies use the same 89-stock universe, base transaction cost, long-only budget convention, and weight-based execution protocol. QQQ is not part of the same-universe fairness comparison; it is included only as an external market reference. The values of $K$, the covariance lookback, and execution hyperparameters are listed in Appendix~\ref{app:implementation_details}.

\paragraph{HRT variants.}
We use the following variants to isolate the contribution of each component. \textbf{HRT-Base} is a hierarchical PPO--TD3 model using market forecasts and sentiment, without the sparse confidence filter or activation penalty, without text-risk scores, and with a return-only LLC reward; its categorical HLC can still output hold decisions. \textbf{+ Sparse HLC} adds sparse high-level activation and the activation penalty. \textbf{+ Text Risk Signal} adds text-derived risk scores to the controller states. \textbf{+ Risk-Aware LLC (HRT)} adds turnover, drawdown, and text-risk exposure penalties to the LLC reward. \textbf{HRT w/o Sparse HLC} removes sparse high-level activation from the full HRT model.

\begin{figure*}[t]
\centering
\includegraphics[width=0.92\textwidth]{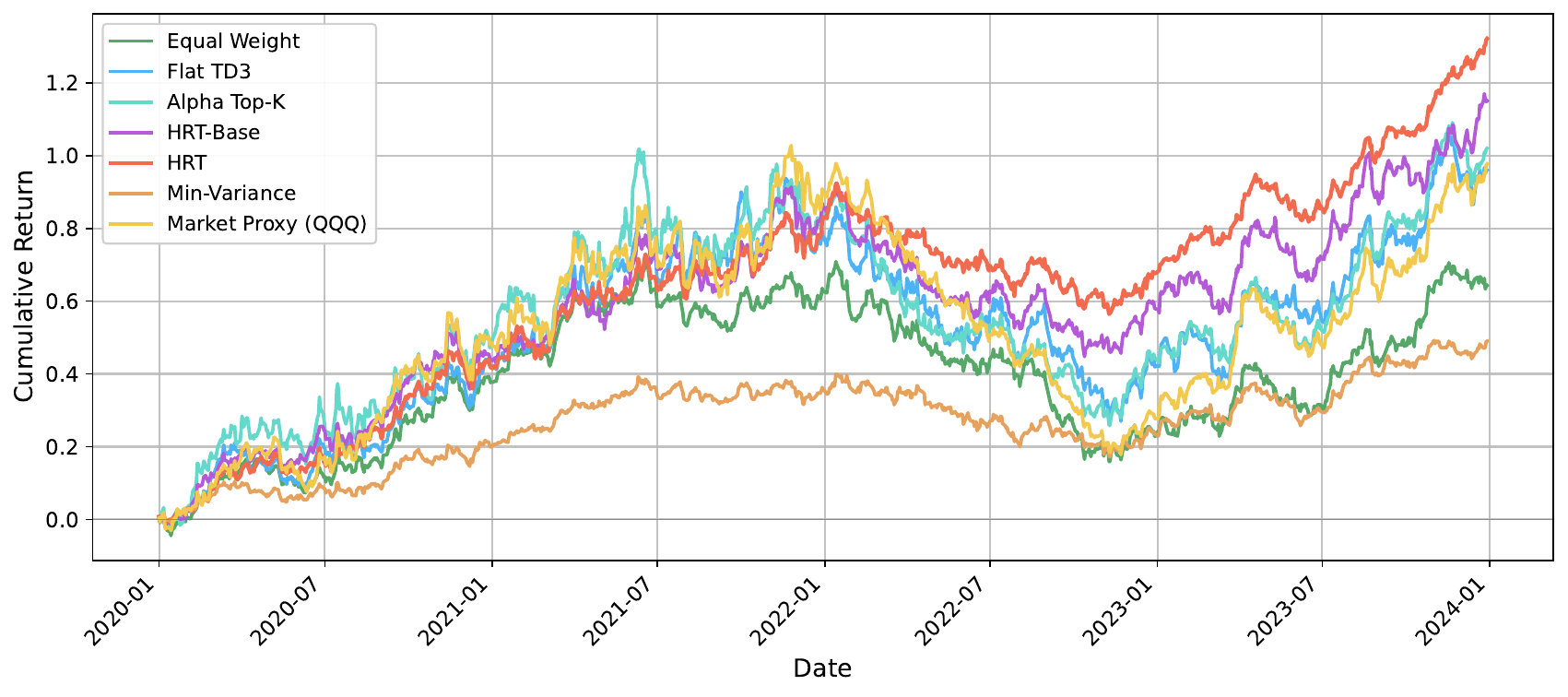}
\caption{Cumulative returns over the 2020--2023 test period. Market Proxy denotes QQQ, an external Nasdaq-100 proxy. HRT does not dominate every year, but it reduces downside during the 2022 drawdown and finishes with the strongest return--risk--cost profile among learning-based strategies.}
\Description{Line chart comparing cumulative returns from 2020 to 2023 for Market Proxy, Equal Weight, Min-Variance, Alpha Top-K, Flat TD3, HRT-Base, and HRT. HRT finishes with the highest cumulative return among learning-based strategies.}
\label{fig:cumulative_return}
\end{figure*}

\paragraph{Metrics and compute.}
Cumulative return is computed as $\prod_t (1+r_t)-1$. We report annualized return as the compound annual growth rate over the test period, $(1+\mathrm{Cum.\ Ret.})^{1/T}-1$. Annualized volatility is $\sqrt{252}\sigma_d$, and Sharpe is the standard daily-return Sharpe, $\sqrt{252}(\bar r_d-r_{f,d})/\sigma_d$ with $r_f=0$. Daily turnover is traded notional divided by portfolio value, $\sum_i |\Delta w^{exec}_{i,t}|$; transaction cost is turnover multiplied by the per-notional cost rate, so no one-half turnover convention is used. We also report maximum drawdown and 5\% CVaR. HLC is trained with PPO and LLC with TD3. Both controllers use two-layer MLPs with 256 hidden units and $5\times 10^5$ training steps. We use two NVIDIA A100 GPUs only for parallelizing seeds and ablations; each training run fits on a single GPU and takes approximately 3--6 hours. All text-derived signals are precomputed and cached, and no LLM fine-tuning or online LLM querying is performed during RL training. Full hyperparameters are listed in Appendix~\ref{app:implementation_details}.

\subsection{Overall Out-of-Sample Performance}
\label{sec:main_performance}

Figure~\ref{fig:cumulative_return} shows cumulative returns over the 2020--2023 test period. HRT participates in the 2020--2021 and 2023 upward regimes, while reducing downside during the 2022 drawdown. Table~\ref{tab:main_performance} reports the corresponding return, risk, and execution metrics.

\begin{table*}[t]
\centering
\caption{Main out-of-sample performance over 2020--2023. Learning-based methods report mean $\pm$ seed standard deviation over five seeds. HRT is not best on every individual risk metric; its advantage is the overall return--risk--cost trade-off among learning-based strategies.}
\label{tab:main_performance}
\small
\resizebox{\textwidth}{!}{%
\begin{tabular}{lrrrrrrr}
\toprule
\textbf{Model} & \textbf{Cum. Ret.} & \textbf{Ann. Ret.} & \textbf{Ann. Vol.} & \textbf{Sharpe} & \textbf{Max DD} & \textbf{CVaR 5\%} & \textbf{Turnover} \\
\midrule
\multicolumn{8}{l}{\emph{External market proxy}} \\
Market Proxy (QQQ) & 0.977 & 0.186 & 0.258 & 0.80 & -0.356 & -0.031 & -- \\
\midrule
\multicolumn{8}{l}{\emph{Same-universe non-RL baselines}} \\
Equal Weight       & 0.643 & 0.132 & 0.215 & 0.70 & -0.292 & -0.027 & 0.015 \\
Min-Variance       & 0.490 & 0.105 & 0.155 & 0.74 & \textbf{-0.205} & \textbf{-0.020} & 0.045 \\
Momentum Top-$K$   & 0.920 & 0.177 & 0.295 & 0.72 & -0.410 & -0.037 & 0.160 \\
Alpha Top-$K$      & 1.020 & 0.192 & 0.285 & 0.82 & -0.340 & -0.033 & 0.210 \\
\midrule
\multicolumn{8}{l}{\emph{Learning-based strategies}} \\
Flat TD3           & $0.960\pm0.110$ & $0.183\pm0.017$ & $0.255\pm0.018$ & $0.83\pm0.11$ & $-0.315\pm0.040$ & $-0.030\pm0.003$ & $0.195\pm0.025$ \\
HRT-Base           & $1.151\pm0.080$ & $0.211\pm0.011$ & $0.220\pm0.010$ & $1.06\pm0.07$ & $-0.305\pm0.025$ & $-0.027\pm0.002$ & $0.112\pm0.010$ \\
\textbf{HRT}       & $\mathbf{1.321\pm0.060}$ & $\mathbf{0.234\pm0.009}$ & $0.208\pm0.010$ & $\mathbf{1.24\pm0.06}$ & $-0.245\pm0.015$ & $-0.028\pm0.003$ & $\mathbf{0.090\pm0.007}$ \\
\bottomrule
\end{tabular}%
}
\end{table*}

HRT achieves the strongest overall return--risk--cost trade-off among the learning-based strategies. Compared with HRT-Base, HRT improves the standard daily-return Sharpe from 1.06 to 1.24 and reduces daily turnover from 0.112 to 0.090, while also improving maximum drawdown from -0.305 to -0.245. Compared with Alpha Top-$K$, HRT obtains higher cumulative return with lower drawdown and substantially lower turnover, suggesting that the hierarchy adds value beyond the supervised alpha signal by translating it into more cost- and risk-controlled portfolio actions. HRT does not dominate every individual risk metric: Min-Variance has lower maximum drawdown and CVaR, and HRT-Base has slightly better one-day CVaR. However, both alternatives earn lower cumulative return and Sharpe than HRT. This pattern is consistent with selecting HRT for the overall return--risk--cost balance rather than for pure tail-risk minimization.

\subsection{Component Ablation}
\label{sec:ablation}

Table~\ref{tab:ablation} isolates the effect of the main design choices. The first four rows add components incrementally, moving from the hierarchical base model to the full HRT system. The final row removes sparsity from the full HRT model as a diagnostic check: it obtains the highest raw cumulative return, but at the cost of higher turnover and drawdown.

\begin{table*}[t]
\centering
\caption{Ablation study. The first four rows add components incrementally. The last row removes sparse activation from the full model to show the cost of unconstrained trading activity. Values are mean $\pm$ seed standard deviation over five seeds. Text-risk exposure is computed ex post for all variants using the same normalized risk scores.}
\label{tab:ablation}
\small
\resizebox{\textwidth}{!}{%
\begin{tabular}{lrrrrr}
\toprule
\textbf{Variant} & \textbf{Cum. Ret.} & \textbf{Sharpe} & \textbf{Max DD} & \textbf{Turnover} & \textbf{Text-Risk Exp.} \\
\midrule
HRT-Base                    & $1.151\pm0.080$ & $1.06\pm0.07$ & $-0.305\pm0.025$ & $0.112\pm0.010$ & $0.43\pm0.02$ \\
+ Sparse HLC                & $1.120\pm0.075$ & $1.10\pm0.06$ & $-0.270\pm0.020$ & $0.101\pm0.008$ & $0.42\pm0.02$ \\
+ Text Risk Signal          & $1.205\pm0.065$ & $1.17\pm0.06$ & $-0.255\pm0.018$ & $0.098\pm0.008$ & $0.39\pm0.02$ \\
+ Risk-Aware LLC (\textbf{HRT}) & $\mathbf{1.321\pm0.060}$ & $\mathbf{1.24\pm0.06}$ & $\mathbf{-0.245\pm0.015}$ & $\mathbf{0.090\pm0.007}$ & $\mathbf{0.35\pm0.02}$ \\
HRT w/o Sparse HLC          & $1.365\pm0.095$ & $1.12\pm0.08$ & $-0.335\pm0.035$ & $0.150\pm0.018$ & $0.38\pm0.02$ \\
\bottomrule
\end{tabular}%
}
\end{table*}

The ablation suggests that different components improve different parts of the trading pipeline. Sparse high-level activation sacrifices some raw return but tends to improve Sharpe, drawdown, and turnover. Adding text-derived risk information reduces ex-post text-risk exposure and improves path-level downside control. The full risk-aware LLC further improves the return--risk--cost balance by combining higher Sharpe, lower turnover, lower text-risk exposure, and smaller maximum drawdown. The unconstrained variant confirms that higher raw return can be obtained by trading more aggressively, but this comes with substantially worse drawdown and turnover; therefore the full HRT is selected for risk-adjusted and cost-aware performance rather than maximum raw return or pure tail-risk minimization.

\subsection{Execution Robustness}
\label{sec:execution_robustness}

\begin{figure}[t]
\centering
\includegraphics[width=\columnwidth]{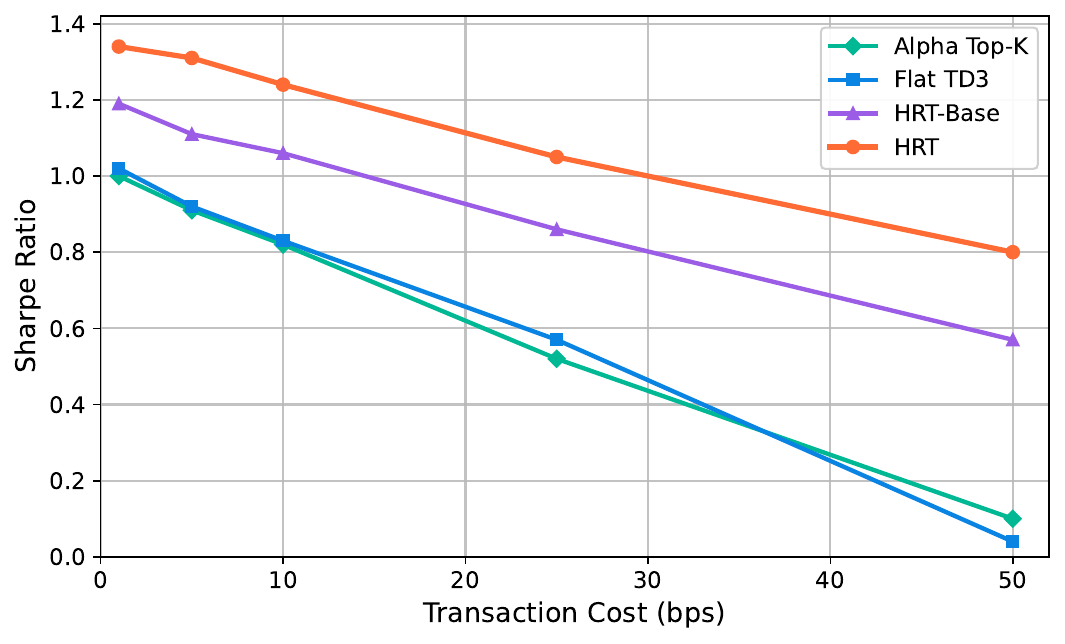}
\caption{Transaction-cost sensitivity. Policies selected under the 10 bps validation setting are replayed under alternative cost assumptions without retraining. All strategies deteriorate as transaction costs increase, but HRT achieves the highest Sharpe ratio among the reported strategies across the evaluated 1--50 bps cost settings. The higher-turnover Alpha Top-$K$ and Flat TD3 baselines degrade more sharply, while HRT and HRT-Base are more resilient due to hierarchical selection and lower trading activity.}
\Description{Line chart showing Sharpe ratio under transaction costs from 1 to 50 basis points for Alpha Top-K, Flat TD3, HRT-Base, and HRT. HRT remains highest among the reported strategies across the evaluated cost settings.}
\label{fig:cost_sensitivity}
\end{figure}

A central motivation for the LLC is that directional signals should not be converted into excessive turnover. We therefore evaluate out-of-sample cost sensitivity by applying the policies selected under the 10 bps validation setting to alternative transaction-cost assumptions without retraining. Figure~\ref{fig:cost_sensitivity} reports standard daily-return Sharpe ratios under costs from 1 to 50 bps.

At the base 10 bps setting, HRT achieves a standard daily-return Sharpe ratio of 1.24. As transaction costs rise, all strategies lose risk-adjusted performance, but the decline is substantially sharper for the higher-turnover Alpha Top-$K$ and Flat TD3 baselines. HRT-Base remains more resilient, and the full HRT achieves the highest Sharpe ratio among the reported strategies across the evaluated 1--50 bps cost settings. Under severe 50 bps cost stress, HRT retains a positive Sharpe ratio of 0.80, compared with 0.57 for HRT-Base and near-zero values for Alpha Top-$K$ and Flat TD3. This pattern supports the role of sparse high-level activation and risk-aware execution in improving cost robustness by limiting unnecessary trading activity. Additional trading behavior statistics, including active names, annualized cost drag, concentration, and text-risk exposure, are reported in Appendix~\ref{app:additional_results}.

\section{Discussion and Conclusion}
\label{sec:conclusions}
This paper presents the Hierarchical Reinforced Trader (HRT), a bi-level reinforcement learning framework that decomposes equity trading into sparse directional selection and risk-aware portfolio execution. The High-Level Controller produces stock-level increase, reduce, or hold decisions from compact market and text-derived signals, while the Low-Level Controller converts these decisions into feasible portfolio adjustments under cost, turnover, drawdown, and text-risk considerations. Empirically, HRT achieves the strongest overall return--risk--cost trade-off among learning-based strategies on an open stock-news benchmark. Its gains are not driven by optimizing a single metric in isolation: HRT improves risk-adjusted return, lowers turnover, reduces maximum drawdown relative to the hierarchical base model, and remains more robust under transaction-cost stress. These results support the central premise of the paper: in text-aware portfolio control, separating what to trade from how aggressively to trade can make financial RL more selective, auditable, and execution-aware.

The framework also has limitations. Our evaluation follows the filtered 89-stock universe and 2020--2023 horizon supported by the public stock-news benchmark; this fixed-universe protocol evaluates the proposed hierarchical execution mechanism rather than claiming production-level live-trading performance. The backtest uses weight-based fractional execution without integer-share rounding, market impact, or capacity constraints. HRT is signal-agnostic, but its performance still depends on the quality, calibration, and timestamp alignment of upstream market and text-derived signals. The factorized HLC is also a scalable approximation and does not fully model all cross-asset dependencies. Future work should therefore evaluate HRT on point-in-time dynamic universes, enrich the structured text interface with event relevance and confidence scores, and incorporate integer-share rounding, liquidity, market impact, and capacity constraints into the LLC.

Overall, HRT provides a practical template for building financial RL systems that are not only return-seeking, but also selective and execution-aware. Rather than asking a single flat policy to jointly learn stock selection, position sizing, cost control, and risk management, HRT assigns these responsibilities to a factorized sparse high-level controller and a risk-aware low-level controller. This design allows market forecasts and benchmark-provided text-derived sentiment and risk signals to enter the decision process as structured, auditable inputs, while the final trading behavior remains constrained by turnover, drawdown, and execution feasibility. Under a fixed public benchmark protocol, HRT improves the learning-based return--risk--cost trade-off and offers a reusable foundation for future research on text-aware portfolio control, hierarchical trading agents, and risk-aware financial decision-making.

\bibliographystyle{ACM-Reference-Format}
\bibliography{references}

@misc{benhenda2025finrldeepseek,
  author  = {Benhenda, Mostapha},
  title   = {{FinRL-DeepSeek}: {LLM}-Infused Risk-Sensitive Reinforcement Learning for Trading Agents},
  howpublished = {arXiv preprint arXiv:2502.07393},
  year    = {2025}
}

@book{bertsekas2012dynamic,
  author    = {Bertsekas, Dimitri P.},
  title     = {Dynamic Programming and Optimal Control: Volume I},
  volume    = {4},
  publisher = {Athena Scientific},
  address   = {Nashua, NH, USA},
  year      = {2012}
}

@misc{dong2024fnspid,
  author  = {Dong, Zihan and Fan, Xinyu and Peng, Zhiyuan},
  title   = {{FNSPID}: A Comprehensive Financial News Dataset in Time Series},
  howpublished = {arXiv preprint arXiv:2402.06698},
  year    = {2024}
}

@inproceedings{fujimoto2018addressing,
  author       = {Fujimoto, Scott and van Hoof, Herke and Meger, David},
  title        = {Addressing Function Approximation Error in Actor-Critic Methods},
  booktitle    = {Proceedings of the 35th International Conference on Machine Learning},
  pages        = {1587--1596},
  publisher    = {PMLR},
  address      = {Stockholm, Sweden},
  year         = {2018}
}

@inproceedings{han2023select,
  author    = {Han, Weiguang and Zhang, Boyi and Xie, Qianqian and Peng, Min and Lai, Yanzhao and Huang, Jimin},
  title     = {Select and Trade: Towards Unified Pair Trading with Hierarchical Reinforcement Learning},
  booktitle = {Proceedings of the 29th ACM SIGKDD Conference on Knowledge Discovery and Data Mining},
  pages     = {4123--4134},
  publisher = {Association for Computing Machinery},
  address   = {New York, NY, USA},
  year      = {2023}
}

@inproceedings{ke2017lightgbm,
  author    = {Ke, Guolin and Meng, Qi and Finley, Thomas and Wang, Taifeng and Chen, Wei and Ma, Weidong and Ye, Qiwei and Liu, Tie-Yan},
  title     = {{LightGBM}: A Highly Efficient Gradient Boosting Decision Tree},
  booktitle = {Advances in Neural Information Processing Systems},
  volume    = {30},
  pages     = {3146--3154},
  publisher = {Curran Associates, Inc.},
  address   = {Red Hook, NY, USA},
  year      = {2017}
}

@article{koratamaddi2021market,
  author  = {Koratamaddi, Prahlad and Wadhwani, Karan and Gupta, Mridul and Sanjeevi, Sriram G.},
  title   = {Market Sentiment-Aware Deep Reinforcement Learning Approach for Stock Portfolio Allocation},
  journal = {Engineering Science and Technology, an International Journal},
  volume  = {24},
  number  = {4},
  pages   = {848--859},
  year    = {2021}
}

@misc{lillicrap2015continuous,
  author  = {Lillicrap, Timothy P. and Hunt, Jonathan J. and Pritzel, Alexander and Heess, Nicolas and Erez, Tom and Tassa, Yuval and Silver, David and Wierstra, Daan},
  title   = {Continuous Control with Deep Reinforcement Learning},
  howpublished = {arXiv preprint arXiv:1509.02971},
  year    = {2015}
}

@misc{liu2018practical,
  author  = {Liu, Xiao-Yang and Xiong, Zhuoran and Zhong, Shan and Yang, Hongyang and Walid, Anwar},
  title   = {Practical Deep Reinforcement Learning Approach for Stock Trading},
  howpublished = {arXiv preprint arXiv:1811.07522},
  year    = {2018}
}

@inproceedings{liu2021finrl,
  author    = {Liu, Xiao-Yang and Yang, Hongyang and Gao, Jiechao and Wang, Christina Dan},
  title     = {{FinRL}: Deep Reinforcement Learning Framework to Automate Trading in Quantitative Finance},
  booktitle = {Proceedings of the Second ACM International Conference on AI in Finance},
  pages     = {1--9},
  publisher = {Association for Computing Machinery},
  address   = {New York, NY, USA},
  year      = {2021}
}

@inproceedings{liu2022finrlmeta,
  author    = {Liu, Xiao-Yang and Xia, Ziyi and Rui, Jingyang and Gao, Jiechao and Yang, Hongyang and Zhu, Ming and Wang, Christina Dan and Wang, Zhaoran and Guo, Jian},
  title     = {{FinRL-Meta}: Market Environments and Benchmarks for Data-Driven Financial Reinforcement Learning},
  booktitle = {Advances in Neural Information Processing Systems},
  volume    = {35},
  numpages  = {15},
  publisher = {Curran Associates, Inc.},
  address   = {Red Hook, NY, USA},
  year      = {2022}
}

@article{liu2024dynamic,
  author  = {Liu, Xiao-Yang and Xia, Ziyi and Yang, Hongyang and Gao, Jiechao and Zha, Daochen and Zhu, Ming and Wang, Christina Dan and Wang, Zhaoran and Guo, Jian},
  title   = {Dynamic Datasets and Market Environments for Financial Reinforcement Learning},
  journal = {Machine Learning},
  volume  = {113},
  number  = {5},
  pages   = {2795--2839},
  doi     = {10.1007/s10994-023-06511-w},
  year    = {2024}
}

@article{markowitz1952portfolio,
  author  = {Markowitz, Harry},
  title   = {Portfolio Selection},
  journal = {The Journal of Finance},
  volume  = {7},
  number  = {1},
  pages   = {77--91},
  year    = {1952}
}

@inproceedings{mnih2016asynchronous,
  author    = {Mnih, Volodymyr and Badia, Adria Puigdomenech and Mirza, Mehdi and Graves, Alex and Lillicrap, Timothy and Harley, Tim and Silver, David and Kavukcuoglu, Koray},
  title     = {Asynchronous Methods for Deep Reinforcement Learning},
  booktitle = {Proceedings of the 33rd International Conference on Machine Learning},
  pages     = {1928--1937},
  publisher = {PMLR},
  address   = {New York, NY, USA},
  year      = {2016}
}

@inproceedings{nan2022sentiment,
  author       = {Nan, Abhishek and Perumal, Anandh and Zaiane, Osmar R.},
  title        = {Sentiment and Knowledge Based Algorithmic Trading with Deep Reinforcement Learning},
  booktitle    = {International Conference on Database and Expert Systems Applications},
  pages        = {167--180},
  publisher    = {Springer},
  address      = {Cham, Switzerland},
  year         = {2022}
}

@article{pateria2021hierarchical,
  author  = {Pateria, Shubham and Subagdja, Budhitama and Tan, Ah-Hwee and Quek, Chai},
  title   = {Hierarchical Reinforcement Learning: A Comprehensive Survey},
  journal = {ACM Computing Surveys},
  volume  = {54},
  number  = {5},
  pages   = {1--35},
  year    = {2021}
}

@inproceedings{qin2024earnhft,
  author    = {Qin, Molei and Sun, Shuo and Zhang, Wentao and Xia, Haochong and Wang, Xinrun and An, Bo},
  title     = {{EarnHFT}: Efficient Hierarchical Reinforcement Learning for High Frequency Trading},
  booktitle = {Proceedings of the AAAI Conference on Artificial Intelligence},
  volume    = {38},
  pages     = {14669--14676},
  publisher = {AAAI Press},
  address   = {Palo Alto, CA, USA},
  year      = {2024}
}

@misc{schulman2017proximal,
  author  = {Schulman, John and Wolski, Filip and Dhariwal, Prafulla and Radford, Alec and Klimov, Oleg},
  title   = {Proximal Policy Optimization Algorithms},
  howpublished = {arXiv preprint arXiv:1707.06347},
  year    = {2017}
}

@misc{wang2020deep,
  author  = {Wang, Rundong and Wei, Hongxin and An, Bo and Feng, Zhouyan and Yao, Jun},
  title   = {Deep Stock Trading: A Hierarchical Reinforcement Learning Framework for Portfolio Optimization and Order Execution},
  howpublished = {arXiv preprint arXiv:2012.12620},
  year    = {2020}
}

@inproceedings{yang2020deep,
  author    = {Yang, Hongyang and Liu, Xiao-Yang and Zhong, Shan and Walid, Anwar},
  title     = {Deep Reinforcement Learning for Automated Stock Trading: An Ensemble Strategy},
  booktitle = {Proceedings of the First ACM International Conference on AI in Finance},
  pages     = {1--8},
  publisher = {Association for Computing Machinery},
  address   = {New York, NY, USA},
  year      = {2020}
}

@misc{yang2020qlib,
  author  = {Yang, Xiao and Liu, Weiqing and Zhou, Dong and Bian, Jiang and Liu, Tie-Yan},
  title   = {{Qlib}: An AI-Oriented Quantitative Investment Platform},
  howpublished = {arXiv preprint arXiv:2009.11189},
  year    = {2020}
}

@misc{zhang2023instruct,
  author  = {Zhang, Boyu and Yang, Hongyang and Liu, Xiao-Yang},
  title   = {{Instruct-FinGPT}: Financial Sentiment Analysis by Instruction Tuning of General-Purpose Large Language Models},
  howpublished = {arXiv preprint arXiv:2306.12659},
  year    = {2023}
}

@misc{zhang2024finagent,
  author  = {Zhang, Wentao and Zhao, Lingxuan and Xia, Haochong and Sun, Shuo and Sun, Jiaze and Qin, Molei and Li, Xinyi and Zhao, Yuqing and Zhao, Yilei and Cai, Xinyu and Zheng, Longtao and Wang, Xinrun and An, Bo},
  title   = {A Multimodal Foundation Agent for Financial Trading: Tool-Augmented, Diversified, and Generalist},
  howpublished = {arXiv preprint arXiv:2402.18485},
  year    = {2024}
}

\appendix

\section{Implementation and Hyperparameters}
\label{app:implementation_details}

\begin{table*}[!t]
\centering
\caption{Implementation configuration. Hyperparameters that affect model selection are chosen using the 2019 validation period only.}
\label{tab:implementation_appendix}
\small
\begin{tabular}{p{0.28\textwidth}p{0.62\textwidth}}
\toprule
\textbf{Component} & \textbf{Setting} \\
\midrule
HLC algorithm & PPO \\
LLC algorithm & TD3 \\
Market encoder & LightGBM on OHLCV and technical factors \\
Forecast target & Return from next open execution to next scheduled rebalance \\
Risk proxy & 20-day rolling realized volatility \\
Top-$K$ & $K=30$ for Momentum Top-$K$ and Alpha Top-$K$ \\
Min-Variance lookback & 252 trading days \\
Single-name cap & $w_{\max}=5\%$ \\
LLC turnover budget & $\tau_{\max}=0.20$ \\
Sparse confidence threshold & 0.58, selected on validation \\
Base transaction cost & 10 bps per traded notional \\
Reward penalties & $\lambda_{turn}=0.10$, $\lambda_{dd}=0.05$, $\lambda_{risk}=0.03$, $\lambda_{act}=0.01$ \\
HLC reward schedule & $\alpha_t=\alpha_0\exp(-\lambda t)$, $\alpha_0=1.0$, $\lambda=3\times10^{-6}$ \\
HLC / LLC warm-up steps & $5\times10^4$ / $1\times10^5$ \\
Alternating refinement & One HLC update epoch every five LLC update epochs \\
PPO learning rate & $3\times10^{-4}$ \\
PPO rollout length / epochs & 2048 / 10 \\
PPO clip ratio / entropy coefficient & 0.20 / 0.01 \\
TD3 actor / critic learning rates & $3\times10^{-4}$ / $1\times10^{-3}$ \\
TD3 policy delay & 2 \\
TD3 target smoothing noise & 0.20, clipped at 0.50 \\
TD3 soft update & 0.005 \\
Replay buffer / batch size & $2\times10^5$ / 256 \\
Discount factor & 0.99 \\
Network architecture & Two-layer MLP, 256 hidden units \\
Training steps & $5\times10^5$ \\
Execution assumption & Weight-based next-open execution; no integer-share rounding \\
Model selection & Validation-only selection on 2019 \\
LLM fine-tuning / online querying & None / none \\
\bottomrule
\end{tabular}
\end{table*}

The feasibility layer in Eq.~\eqref{eq:llc_action_chain} is implemented as deterministic clipping and scaling. Hold assets are assigned zero adjustment. Reduce assets can only decrease down to zero weight, and increase assets can only use available cash from existing cash and reductions. Desired increases are clipped to the single-name cap and, when buy demand exceeds available cash or the daily turnover budget, positive active adjustments are scaled proportionally. The procedure never applies a final proportional rescaling to all holdings, so hold assets remain unchanged and active adjustments preserve the HLC direction. The replay buffer stores the resulting executed adjustment. Target actions in TD3 are passed through the same feasibility layer, and the layer is treated as deterministic post-processing during actor updates.

\section{Additional Results}
\label{app:additional_results}

Table~\ref{tab:trading_behavior_appendix} summarizes trading behavior over the 2020--2023 test period. HRT trades about 25\% of the universe on an average day and reduces annualized cost drag relative to HRT-Base, while maintaining lower text-risk exposure. It is not the least concentrated strategy by HHI, which is consistent with the claim that HRT improves the overall return--risk--cost balance rather than optimizing every metric independently.

\begin{table*}[!t]
\centering
\caption{Trading behavior over the 2020--2023 test period. Active ratio is active stocks divided by 89. Turnover is $\sum_i |\Delta w^{exec}_{i,t}|$ without the one-half convention. Annualized cost drag is daily turnover $\times$ 10 bps $\times$ 252. HHI is $\sum_i w_{i,t}^2$, averaged over the test period. Text-risk exposure is $\sum_i w_{i,t}\rho_{i,t}$ using the normalized and clipped text-risk score; it is computed ex post for all strategies.}
\label{tab:trading_behavior_appendix}
\small
\begin{tabular*}{\textwidth}{@{\extracolsep{\fill}}lrrrrrr}
\toprule
\textbf{Model} 
& \shortstack{\textbf{Active}\\\textbf{Stocks}} 
& \shortstack{\textbf{Active}\\\textbf{Ratio}} 
& \textbf{Turnover} 
& \shortstack{\textbf{Ann. Cost}\\\textbf{Drag}} 
& \textbf{HHI} 
& \shortstack{\textbf{Text-Risk}\\\textbf{Exp.}} \\
\midrule
Alpha Top-$K$       & 30 & 0.34 & 0.210 & 0.053 & 0.033 & 0.48 \\
Flat TD3            & 40 & 0.45 & 0.195 & 0.049 & 0.050 & 0.50 \\
HRT-Base            & 27 & 0.30 & 0.112 & 0.028 & 0.044 & 0.43 \\
\textbf{HRT}        & \textbf{22} & \textbf{0.25} & \textbf{0.090} & \textbf{0.023} & 0.048 & \textbf{0.35} \\
HRT w/o Sparse HLC  & 38 & 0.43 & 0.150 & 0.038 & 0.041 & 0.38 \\
\bottomrule
\end{tabular*}
\end{table*}

Table~\ref{tab:yearly_appendix} reports yearly robustness results. HRT does not dominate the market proxy in every bull-market year. Its advantage is clearest in the 2022 drawdown, where it reduces downside relative to HRT-Base while still participating in the 2023 recovery.

\begin{table*}[!t]
\centering
\caption{Yearly robustness analysis. HRT is not designed to dominate the market proxy in every bull-market year; its advantage is most visible in downside control during adverse regimes.}
\label{tab:yearly_appendix}
\small
\begin{tabular*}{\textwidth}{@{\extracolsep{\fill}}llrrrrr}
\toprule
\textbf{Year} 
& \textbf{Regime} 
& \shortstack{\textbf{Market Proxy}\\\textbf{Ret.}} 
& \shortstack{\textbf{HRT-Base}\\\textbf{Ret.}} 
& \shortstack{\textbf{HRT}\\\textbf{Ret.}} 
& \shortstack{\textbf{HRT-Base}\\\textbf{Max DD}} 
& \shortstack{\textbf{HRT}\\\textbf{Max DD}} \\
\midrule
2020 & COVID shock / recovery       & 0.486  & 0.440  & 0.420  & -0.285 & -0.245 \\
2021 & Bull market                  & 0.274  & 0.290  & 0.290  & -0.110 & -0.100 \\
2022 & Fed tightening / bear market & -0.326 & -0.150 & -0.085 & -0.305 & -0.235 \\
2023 & Tech-led recovery            & 0.549  & 0.362  & 0.385  & -0.130 & -0.115 \\
\bottomrule
\end{tabular*}
\end{table*}

\clearpage

\end{document}